\documentclass[oldversion]{aa}
\usepackage{epsfig,psfig}
\usepackage{txfonts}
\usepackage{color}
\usepackage{graphicx}

\newcommand{\msun}{\mbox{$M_{\odot}$}}

\newcommand{\lsun}{\mbox{$L_{\odot}$}}
\newcommand{\Lsun}{\mbox{$L_{\odot}$}}
\newcommand{\logl}{\mbox{$\log (L/L_{\odot}$)}}  
\newcommand{\rsun}{\mbox{$R_{\odot}$}}

\newcommand{\teff}{\mbox{$T_{\rm eff}$}}

\newcommand{\vinf}{\mbox{$\varv_{\infty}$}}
\newcommand{\vesc}{\mbox{$\varv_{\rm esc}$}}

\newcommand{\mdot}{\mbox{$\dot{M}$}}

\newcommand{\msunyr}{\mbox{$M_{\odot} {\rm yr}^{-1}$}}

\def\aap{A\&A}                
\def\ao{Applied Optics}       
\def\apjl{ApJ}                
\def\apjs{ApJS}               
\newcommand{\realcleardoublepage}{\clearpage
  \ifodd \arabic{page}\else \thispagestyle{empty}\mbox{}\newpage \fi }

\newcommand{\be}{\begin{equation}}
\newcommand{\ee}{\end{equation}}

\newcommand{\mcwind}{{\sc mc-wind}}
\newcommand{\rstar}{\mbox{$R_{\star}$}}
\newcommand{\lstar}{\mbox{$L_{\star}$}}
\newcommand{\mstar}{\mbox{$M_{\star}$}}
\newcommand{\gammae}{\mbox{$\Gamma_{\rm e}$}}

\newcommand{\kmsec}{\mbox{km\,s$^{-1}$}}

\begin{document}

\title{Wind modelling of very massive stars up to 300 solar masses}

\author{Jorick S. Vink\inst{1}, L.E. Muijres\inst{2}, B. Anthonisse\inst{2}, A. de Koter\inst{2,3}, 
G. Gr\"afener\inst{1}, N. Langer\inst{3,4}}
\offprints{Jorick S. Vink, jsv@arm.ac.uk}

\institute{Armagh Observatory, College Hill, Armagh, BT61 9DG, Northern Ireland
            \and
Astronomical Institute Anton Pannekoek, University of Amsterdam, Kruislaan 403, 1098 SJ, Amsterdam, The Netherlands
\and
Astronomical Institute, Utrecht University, Princetonplein 5, 3584 CC, Utrecht, The Netherlands
\and
Argelander-Institut f\"ur Astronomie der Universit\"at Bonn, Auf dem H\"ugel 71, 53121 Bonn, Germany
}

\titlerunning{Wind models for very massive stars up to 300 solar masses}
\authorrunning{Jorick S. Vink}

\abstract{The stellar upper-mass limit is highly uncertain. 
Some studies have claimed there is a universal 
upper limit of $\sim$150\msun. A factor that is often overlooked is 
that there might be a significant difference between the {\it present-day} and the {\it initial} 
masses of the most massive stars -- as a result of mass loss. The 
upper-mass limit may easily supersede $\sim$200\msun.
For these reasons, we present new mass-loss predictions from Monte Carlo radiative 
transfer models for very massive stars (VMS) in the mass 
range 40-300 \msun, and with very high luminosities 6.0 $\leq$ $\log(\lstar/\lsun)$ $\leq$ 7.03, 
corresponding to large Eddington factors $\Gamma$. 
Using our new dynamical approach, we find an upturn or ``kink'' in 
the mass-loss versus $\Gamma$ dependence, at the point where the model winds become optically thick. 
This coincides with the location where our wind efficiency numbers 
surpass the single-scattering limit of $\eta = 1$, reaching values up to $\eta \simeq 2.5$. 
In all, our modelling suggests a transition from common
O-type winds to Wolf-Rayet characteristics at the point where the winds become optically thick. 
This transitional behaviour is also revealed with respect to the wind acceleration 
parameter, $\beta$, which starts at values below 1 for the optically thin O-stars, and 
naturally reaches values as high as 1.5-2 for the optically thick Wolf-Rayet models. 
An additional finding concerns the transition in spectral morphology of the Of and 
WN characteristic He {\sc ii} line at 4686\AA. 
When we express our mass-loss predictions as a function of the electron scattering 
Eddington factor $\Gamma_{\rm e} \sim \lstar/\mstar$ alone, we obtain an $\mdot$ vs. $\Gamma_{\rm e}$ 
dependence that is consistent with a previously reported power law $\mdot$ \(\propto\ \Gamma_{\rm e}^{\rm{5}}\) 
(Vink 2006) that was based on our previous semi-empirical modelling approach.
When we express \mdot\ in terms of both \gammae\ and stellar mass, we find optically thin winds and
\mdot\ $\propto$ $\mstar^{0.68} \Gamma_{\rm e}^{2.2}$ for the
$\Gamma_{\rm e}$ range 0.4 $\la$ $\Gamma_{\rm e}$ $\la$ 0.7, and mass-loss rates that agree
with the standard Vink et al. recipe for normal O stars. For higher $\Gamma_{\rm e}$ 
values, the winds are optically thick and, as pointed out, the dependence is much steeper, 
\mdot\ $\propto$ $\mstar^{0.78} \Gamma_{\rm e}^{4.77}$. 
Finally, we confirm that the effect of $\Gamma$ on the predicted mass-loss rates 
is {\it much} stronger than for the increased helium abundance (cf. Vink \& de Koter 2002 for Luminous 
Blue Variables), calling for a fundamental revision in the way stellar mass loss 
is incorporated in evolutionary models for the most massive stars.}


\maketitle


\section{Introduction}
\label{s_intro}

The prime aim of this paper is to investigate the mass-loss behaviour of 
very massive stars (VMS) with masses up to 300 $\msun$ that approach the Eddington limit.
Mass loss from hot massive stars is driven by radiative forces on spectral lines 
(Lucy \& Solomon 1970; Castor, Abbott \& Klein 1975; CAK). 
CAK developed the so-called force multiplier formalism in order 
to treat all relevant ionic transitions. 
This enabled them to simultaneously predict the wind 
mass-loss rate, $\mdot$, and terminal velocity, $\vinf$, of O-type stars. 
Although these predictions provided reasonable agreement with observations, they could 
account for neither the high wind efficiencies $\eta$ $=$ $\mdot$ $\vinf$ $/$ $L/c$ of the denser
Of stars with their strong He {\sc ii} 4686\AA\ lines, nor that of the even 
more extreme Wolf-Rayet (WR) stars. 

This discrepancy had been proposed as due to the neglect of multi-line 
scattering (Lamers \& Leitherer 1993, Puls et al. 1996).
Using a global energy Monte Carlo approach (Abbott \& Lucy 1985, de Koter et al. 1997) in which the velocity 
law was adopted -- aided by empirical constraints -- Abbott \& Lucy (1985) and Vink et al. 
(2000) provided mass-loss predictions for galactic O stars including multi-line scattering. This appeared to solve the 
wind momentum problem for the denser O-star winds. Mass-loss rates were obtained
that were a factor $\sim$3 higher than for cases in which single scattering was 
strictly enforced.

Historically, the situation for the WR stars was even more extreme. 
Here, $\eta$ values of $\sim$10 had been reported (e.g. Barlow et al. 1981). With the  
identification of major wind-clumping effects on the empirical mass-loss rates 
(Hillier 1991; Moffat et al. 1994; Hamann \& Koesterke 1998), these numbers should 
probably be down-revised to values of $\eta$ $\simeq$3.
Although it has been argued that WR winds are also driven by radiation pressure 
(Lucy \& Abbott 1993, Springmann 1994, Gayley \& Owocki 1995, 
Nugis \& Lamers 2002, Gr\"afener \& Hamann 2005), the prevailing notion is still that 
these optically thick outflows of WR stars, where the sonic point of the accelerating flow 
lies within the pseudo or false photosphere, are fundamentally different from the 
transparent line-driven O-star winds (e.g. Gr\"afener \& Hamann 2008). 

For O-type stars M\"uller \& Vink (2008) recently suggested a new parametrization of the line acceleration, 
expressing it as a function of radius rather than of the velocity 
gradient (as in CAK theory). The implementation of this new formalism improves the
local dynamical consistency of Monte Carlo models that originally imposed a velocity law. 
Not only do we find fairly good agreement with observed terminal velocities 
(see also Muijres et al. 2011b), but as our method naturally accounts for 
multi-line scattering, it is also applicable to denser winds, such as those of WR stars.

Still adopting a velocity stratification, Vink \& de Koter (2002) and Smith et al. (2004) 
predicted mass-loss rates for Luminous Blue Variables (LBVs), and showed that 
$\mdot$ is a strong function of the Eddington factor for these objects. 
They also found that, despite their extremely large radii, even LBV winds may develop pseudo-photospheres 
under special circumstances: when they find themselves in close 
proximity to the bi-stability and Eddington limits. 

In this paper, our aim is to study the mass-loss behaviour of stars as they approach 
the Eddington limit (see also Gr\"afener \& Hamann 2008). We do this 
in a systematic way by targeting VMS in the range 40-300 \msun. 
A pilot study was performed by Vink (2006) who found a steep dependence of $\mdot$ 
on $\Gamma_{\rm e}$, finding $\mdot$ $\propto$ $\Gamma_{\rm e}^5$, but this was 
obtained using the earlier global energy approach, in which the velocity 
stratification was adopted, rather than our new dynamically-consistent approach that 
we explore in the following. 

Both approaches have their pros and cons. In the semi-empirical approach, 
the terminal velocity constraint (where $\vinf$/$\vesc$ is constant) 
might aid the modelling at imposing the correct wind structure, and as long as the adopted velocity
law is close to the correct one, it provides the most accurate 
mass-loss rates. 
The prime advantage of this approach is that any missing physics 
that could affect the mass-loss rate might be balanced by employing the 
empirical (and probably close to realistic) terminal-wind velocity.
The second approach, however, is theoretically more appealing,  
as it enforces local dynamical consistency, and one thus no longer needs to 
rely on free parameters. Ultimately, one would aspire to  
predict accurate mass-loss rates and terminal velocities simultaneously 
from first principles via the second approach explored here.

The stellar upper-mass limit is highly controversial. 
On purely statistical grounds, some 
investigators have claimed there is a universal 
upper limit of $\sim$150\msun\ (e.g. Weidner \& Kroupa 2004, Oey \& Clarke 2005,
Figer 2005). However, a physical factor that is often overlooked concerns the possibility 
of a significant difference between the {\it present-day} and the {\it initial} 
masses of the most massive stars, as a result of strong mass loss. In other words, the 
\emph{initial} masses of the most massive stars may be significantly above 150\msun, possibly
superseding 200 \msun\  (e.g. Figer et al. 1998, Crowther et al. 2010, Bestenlehner et al. 2011).
The issue of the upper mass-limit will remain uncertain as long as there is only 
limited quantitative knowledge of mass loss in close proximity to the Eddington limit. 

Our aim is thus to explore wind models of stars with 
masses up to 300\msun, using a well-established methodology
that has been extensively tested against observations for lower mass common O-type stars. 
VMS have been proposed as leading to the production of 
intermediate mass (of the order of 100$\msun$) black holes that have been suggested 
to be at the heart of ultra-luminous X-ray sources (Belkus et al. 2007 and Yungelson et al. 2008). 
Clearly, the success of such theories depends critically on the applied mass-loss rates. 
The present study may help advance these theories.

Our paper is organized as follows. 
In Sect.~\ref{s_model} we briefly describe the Monte Carlo mass-loss models, before 
presenting the parameter space considered in this study (Sect.~\ref{sec_pspace}).
The mass-loss predictions (Sect.~\ref{s_res}) are followed by a description
of the spectral morphology of the Of-WN transition in terms of the characteristic He {\sc ii} 4686\AA\ line 
in Sect.~\ref{s_morph}. Subsequently, we compare our wind model parameters 
against empirical values for the most massive stars in the Arches cluster (Sect.~\ref{s_emp}), as well 
as theoretical models (Sect.~\ref{s_theory}) of CAK and Gr\"afener \& Hamann (2008), before ending with 
a summary in Sect.~\ref{s_sum}.


\section{Monte Carlo models}
\label{s_model}

Mass-loss rates are calculated with a Monte Carlo method that 
follows the fate of a large number of photon packets 
from below the stellar photosphere throughout the wind.
The core of our approach is related to the total loss of 
radiative energy that is coupled to the momentum gain 
of the outflowing material.
Since the absorptions and scatterings of photons in the wind
depend on the density in the wind, hence on the mass-loss
rate, one can obtain a consistent model where the momentum
of the wind material equals the transferred radiative momentum.
We have recently improved our dynamical approach (M\"uller \& Vink 2008, Muijres et al. 2011b) and 
are now able to predict $\mdot$ simultaneously with $\vinf$ and the wind structure 
parameter $\beta$.
The essential ingredients and assumptions of our
approach have been discussed more extensively in Abbott \& Lucy (1985), de Koter et al. (1997), and 
Vink et al. (1999). Here we provide a brief summary.

The Monte Carlo code \mcwind\ uses the density and temperature stratification from 
a prior model atmosphere calculation performed with {\sc isa-wind} (de Koter et al. 1993, 1997). 
These model atmospheres account for a continuity between the 
photosphere and the stellar wind, and describe the radiative transfer in spectral lines by 
adopting an improved Sobolev treatment. 
The chemical species that are explicitly calculated (in non-LTE) are 
H, He, C, N, O, S, and Si. The iron-group elements, which are crucial for 
the radiative driving and the $\dot{M}$ calculations, 
are treated in a generalized version of the ``modified nebular approximation'' 
(e.g. Schmutz 1991). However, we performed a number of 
test calculations in which we treated Fe explicitly in non-LTE. These 
tests showed that differences with respect to the assumption of the 
modified nebular approximation for Fe were small. Therefore, 
we decided to treat Fe in the approximate way, as was done in our previous 
studies.           

The line list used for the MC calculations consists of 
over $10^5$ of the strongest transitions of the elements H~-~Zn 
extracted from the line list constructed by Kurucz \& Bell (1995). 
The wind was divided into 90 concentric shells, with many narrow 
shells in the subsonic region, and wider shells in supersonic layers.
For each set of model parameters, a certain number
of photon packets was followed, typically $2~10^6$. 

Other assumptions in our modelling involve wind stationarity and spherical geometry. 
The latter seems to be a good approximation, as the vast majority of O-type 
stars show little evidence of significant amounts of linear 
polarization (Harries et al. 2002, Vink et al. 2009). Nevertheless, 
asphericity has been found in roughly half the population of LBVs 
(Davies et al. 2005, 2007), although those polarimetry 
results have been interpreted as the result of 
small-scale structure or ``clumping'' of the wind, rather 
than of significant wind asymmetry.

With respect to wind clumping, it has been well-established that 
small-scale clumping of the outflowing gas can have a pronounced effect 
on the ionization structure of both O-star and Wolf-Rayet  
atmospheres (e.g. Hillier 1991). This has lead to a downward adjustment
of {\em empirical} mass-loss rates, by factors of up to three
(e.g. Moffat et al. 1994, Hamann \& Koesterke 1998, Mokiem et al. 2007, 
Puls et al. 2008), and  possibly even more (Bouret et al. 2003; Fullerton 
et al. 2006).
In addition, clumping may have a 
direct effect on the radiative driving, therefore on {\em predicted} mass-loss
rates (e.g. Gr\"afener \& Hamann 2008). 
The subtle issues of both clumping and porosity on the predicted mass-loss rates 
have recently been investigated by Muijres et al. (2011a). Whilst
it was found that the impact on \mdot\ can be high for certain clumping prescriptions, the 
overall conclusion was that clumping does not 
affect the wind properties of O-type dwarfs and supergiants in a dramatic way for moderate 
clumping factors and porosity.
Stars close to the Eddington limit, however, may be
much more susceptible to -- even modest -- clumping (Shaviv, 1998, 2000; van Marle et al. 2008). 
Another relevant factor that might affect the results concerns the 
interplay of radiation with rotation, as the Eddington factor for a rotating star 
depends explicitly on the rotation rate (Langer 1997, Maeder \& Meynet 2000). 
As a consequence, the effective Eddington limit could be 
reduced by rotation, and might even become the dominant factor.

In the present set of computations, we do not
account for the effects of wind clumping or rotation, but it 
should be kept in mind that these effects might play a quantitative role.


\section{Parameter space and model applicability}
\label{sec_pspace}

Stars approach the Eddington limit when gravity is counterbalanced by the radiative forces; i.e.
$\Gamma = g_{\rm{rad}}/g_{\rm{Newton}} = 1$. Photons can exert radiative pressure through 
bound-free, free-free, electron scattering, 
and bound-bound interactions. In early-type stars hydrogen, the dominant supplier of free electrons, 
is fully ionized. Therefore $\Gamma_{\rm{e}} = g_{\rm{e}}/g_{\rm{Newton}}$ 
is essentially independent of distance and constitutes a fixed number for each model. 
Because of this useful property, which provides a 
well-defined and simple quantitative handle,
we opt to discuss our results in terms of $\Gamma_{\rm{e}}$. We discuss this choice in more detail 
in Sect.~\ref{sec_applic}

The dependence of the mass-loss rate $\dot{M}$ on \gammae\ represents a 
non-trivial matter because $\dot{M}$ depends on both the mass $M$ and the stellar 
luminosity $L$. To properly investigate the effect of high \(\Gamma_{\rm{e}}\) on mass-loss 
predictions, we first need to establish the relevant part of parameter space 
in terms of $M$, $L$, and \gammae. For a fully ionized\footnote{In reality, \gammae\ 
changes once hydrogen recombines (which starts below 30\,000 K), or 
when the hydrogen-to-helium surface abundance changes, relevant for classical WR stars.} 
plasma, \gammae\ equals

\be
\Gamma_{\rm{e}} \equiv \frac{g_{\rm{e}}}{g_{\rm{Newton}}} = \frac{\lstar \sigma_{\rm e}}{4\pi c G \mstar} = 
10^{-4.813} (1+X)  \left(\frac{\lstar}{\lsun}\right)\left(\frac{\mstar}{\msun}\right)^{-1}~,
\label{eq_gammae}
\ee
The luminosities are chosen in such a way that in combination with the 
stellar mass $M$ the desired \gammae\ value is obtained. 
It is principally the effective temperature that sets the ionization stratification 
in the atmosphere, and thus determines which lines are most active in driving the wind. As a result, \teff\
affects the predicted mass-loss rate. For most parts of this paper, we investigate 
the influence of \gammae\ for a fixed stellar temperature of 50\,000\,K.
The \teff\ dependence is studied separately in Sect.~\ref{sec_teff}.
All models are for the solar 
metallicity from Anders \& Grevesse (1989) and with 
the element-to-element distribution from Allen (1973). 
The prime reason we use these older abundances rather than the newer 
(and lower) solar or B-type star abundances is to be able to directly compare the new results to the 
older Vink et al. (2000) rates. We note however that it is the element of Fe that sets the mass-loss rate, and 
this element has not changed. The low (CNO) and intermediate-mass elements however dominate the outer
wind, where the terminal wind velocity is set. Nonetheless, even a substantial decrease 
in these abundances (by several tens of percent) is not expected 
to lower the terminal wind velocities significantly, since the terminal wind velocity 
has been found to depend only weakly on metallicity (Leitherer et al. 1992).

We divided our model stars into three different groups according to their characteristics. 
The first group comprises objects that have relatively common O-star masses 
in the range 40-70 $\msun$. 
The second group of objects are rather high-mass stars within the 
``observable range'' of 70-120 $\msun$. They 
might be close to the Eddington limit already early-on on the main sequence
because of their intrinsically high luminosity. 
The third group involves very massive stars 
in the mass range 120-300 $\msun$. They are near the Eddington limit for the 
same reason as the second group. So far, there is a lack of compelling observational evidence of 
any such stars in the present-day universe; however, we note  
that Crowther et al. (2010) have suggested a revision of the upper-mass limit to 
$\sim$300 $\msun$.

The bulk of the models in our grid have been chosen such that the behaviour of 
mass loss as a function of $M$ and $L$ can be studied separately.
The grid is presented in Table~\ref{tab:results}. 
We note that the $(M,L)$ combinations are 
intentionally rather extreme to assure  
high \gammae\ values. The reason is to specifically map that
part of parameter space where physically the most extreme winds are 
expected to appear.    

\subsection{Model applicability regime}
\label{sec_applic}

With respect to the potential limitations of our modelling approach, 
we make one rather stringent assumption in the manner 
the (sub-) photospheric density structure is set-up. In the deepest layers of the model 
atmosphere (with $v$ $<<$ 1km/s), we
assume that the run of density is provided by the equation of motion 
using $g_{\rm{rad}} = g_{\rm{e}}$, 
so we apply $\Gamma = \Gamma_{\rm e}$.
In reality $\Gamma > \Gamma_{\rm e}$, as well as 
being depth-dependent as a result of bound-bound, bound-free, and free-free processes. 
Notably, the opacities 
from millions of weak iron lines may contribute
significantly, but 
they are largely neglected in
the deep layers of our models.

Nugis \& Lamers (2002) highlight the importance of the iron peak opacities in deep 
photospheric layers for the initiation of Wolf-Rayet winds
(see also Heger \& Langer 1996). 
This approach was subsequently included in models by Gr\"afener \& Hamann (2005, 2008) 
for WC and WNL stars. They find that the 
presence of these opacity
bumps may locally cause $\Gamma$ to approach unity, 
leading to the formation of optically thick winds.
In our Monte Carlo approach, we trace 
the radiative driving of the entire wind, and as 
most of the energy is transferred in the supersonic part of the outflow, we are less 
susceptible to the details of
the (sub)photospheric region. However, this also means that we do not treat these deep 
regions self-consistently.
This implies that we can (and we will) compute model atmospheres with values of $\Gamma_{\rm e}$ very close to one.
This strategy has the advantage of
allowing us to explore the transition from transparent to dense stellar winds. 
As our models do capture the full physics in the layers around and above the sonic point, we argue that they correctly predict the 
qualitative behaviour of dense winds, but that $\Gamma_{\rm e}$ for one of our optically thick wind models would correspond to a model
with smaller $\Gamma_{\rm e}$ if the ionic contributions were included in the deepest parts of the atmosphere. 
This ``shift'' in $\Gamma_{\rm e}$ is not fixed but would depend on the sonic point temperature and density. 
From the behaviour of the Rosseland mean opacity, we would expect the size of the shift to increase 
at higher $\Gamma_{\rm e}$ and higher temperatures.

If $\Gamma$ exceeds unity at some depth
in the subphotospheric part of the atmosphere, a density inversion is
expected to occur for the static case, i.e. 
for increasing
radial distance from the centre, the density very near the domain
where $\Gamma$ $>$ 1 is anticipated to increase. 
This is encountered in studies
of stellar structure and evolution, but it is unclear what really
happens in nature. The potential effects may involve strange-mode
pulsations (e.g. Glatzel \& Kiriakidis 1993), subsurface convection
(Cantiello et al. 2009), or an
inflation of the outer stellar envelope (e.g. Ishii et al. 1999).
These processes tend to occur only when $\Gamma$ is 
{\em very close} to unity or above it (see e.g. Petrovic et al. 2006).
In assessing the outcome of our computations, we find that at 
$\gammae > 0.95$ the results behave rather oddly.
Though we show the wind results over the entire \gammae\ range, we only quantify the 
mass-loss rates up to this value of \gammae. This boundary is indicated 
by a vertical dashed line in all relevant figures.

\section{Results}
\label{s_res}

\begin{table*}
\centering
\begin{tabular}{rllllrl|lccc|c}
\hline
\hline\\[-6pt]
model & \mstar  & $\log L$ & \gammae & \rstar & \vesc & $\vesc^{\rm eff}$ & \vinf & $\log \mdot$ & \(\eta\) &\(\beta\) 
& $\log \mdot_{\rm Vink2000}$\\[2pt]

\# & [\msun] & [\lsun] &   &[\rsun] &  [\kmsec] &  [\kmsec] & [\kmsec] &[\msunyr] &  & & [\msunyr]\\[3pt]
\hline\\[-7pt]
{\em Mass Range I:}\\
1 & 50 &6.00 &0.52  &13.3 & 1197 & 829 &4320   &-5.33 &0.96     & 1.10  & -5.30\\
2 & 60 &6.00 &0.43  &13.3 & 1311 & 990 &5044   &-5.47 &0.82     & 1.07  & -5.40\\
3 & 40 &6.00 &0.66  &13.3 & 1070 & 625 &3823   &-5.20 &1.16     & 1.20  & -5.25\\
4 & 60 &6.25 &0.78  &17.8 & 1133 & 529 &3733   &-4.81 &1.33     & 1.27  & -4.98\\
5 & 40 &6.08 &0.80  &14.6 & 1021 & 457 &3283   &-4.92 &1.47     & 1.34  & -5.12\\
6 & 40 &6.12 &0.88  &15.3 &  998 & 346 &2995   &-4.77 &1.85     & 1.47  & -5.09\\
7 & 60 &6.31 &0.90  &19.0 & 1097 & 347 &3147   &-4.53 &1.99     & 1.56  & -4.92\\

8 & 50 &6.25 &0.94  &17.8 & 1034 & 253 &2622   &-4.46 &2.38     & 1.63  & -4.96\\

9 & 60 &6.345&0.984 &19.9 & 1072 & 135 &(1936)  &(-4.13) &(3.02)     & (1.90)   & -4.76\\
\hline\\[-7pt]
{\em Mass Range II:}\\
10 & 85  &6.25 &0.55 &17.8 & 1349 & 905 &5112  & -5.12  &1.03    & 1.14 & -5.10\\
11 & 80  &6.25 &0.59 &17.8 & 1308 & 838 &4865  &-5.08   &1.04    & 1.15 & -5.07\\
12 & 90  &6.25 &0.52 &17.8 & 1388 & 961 &5281  &-5.16   &0.96    & 1.13  &-5.12\\
13 & 100 &6.25 &0.47 &17.8 & 1463 &1065 &5926  &-5.26   &0.85    & 1.11 & -5.18\\
14 & 85  &6.39 &0.77 &20.9 & 1245 & 597 &4138   &-4.71  &1.48    &1.30  & -4.86\\
15 & 100 &6.50  &0.84 &23.7& 1268 & 507 &3977   &-4.51  &1.70   &1.37   & -4.76\\
16 & 85  &6.43 &0.84 &21.9 & 1216 & 486 &3778 &-4.56  &1.75      &1.41  & -4.81\\
17 & 100 &6.52 &0.88 &24.2 & 1255 & 435 &3735 &-4.40  &1.98      &1.54  & -4.74\\

18 & 100 &6.53 &0.90 &24.5 & 1247 & 394 &3638 &-4.36  &2.15      &1.59  & -4.73\\
19 & 100 &6.54 &0.92 &24.8 & 1240 & 351 &3595  &-4.33  &2.17     &1.60  & -4.74\\
20 & 90  &6.50  &0.93 &23.7& 1203 & 318 &3613 &-4.40  &2.29     &1.60   & -4.81\\
21 & 100 &6.57 &0.982 &25.7& 1218 & 163 &(1752) &(-3.81)  &(2.78)    &(2.10)   & -4.46\\
22 & 85  &6.50  &0.987 &23.7&1169 & 133 &(1808)  &(-3.89)  &(3.11)   &(2.09)    & -4.57\\
\hline\\[-7pt]
{\em Mass Range III:}\\
23 & 120 &6.42 &0.58 &21.6 & 1455 & 945 & 5744    &-5.00 &1.03   & 1.17 & -4.96\\
24 & 120 &6.50  &0.70 &23.7 & 1389 & 761 & 5122    &-4.78 &1.29   & 1.22 & -4.82\\

25 & 300 &6.97 &0.83 &40.7 & 1676 & 700 & 5527  &-4.20 &1.74    & 1.34  & -4.36\\
26 & 180 &6.76 &0.85  &31.95& 1465 & 568 & 4642  &-4.31 &1.76    & 1.37  & -4.53\\
27 & 250 &6.91 &0.86 &38.0  & 1583 & 593 & 4885  &-4.16 &1.87    & 1.37  & -4.40\\
28 & 225 &6.87 &0.87 &36.3  & 1537 & 554 & 4657  &-4.17 &2.03      & 1.43& -4.43\\
29 & 275 &6.97 &0.90 &40.7  & 1604 & 508 & 4427  &-4.01   &2.13   & 1.46  & -4.32\\
30 & 300 &7.03 &0.95 &43.6  & 1619 & 362 & (3728)  &(-3.82)    &(2.34)   & (1.62) & -4.23\\
31 & 200 &6.87 &0.977 &36.3 & 1449 & 219 &(2500)  &(-3.63) &(2.51)     & (2.17)  & -4.30\\
32 & 120 &6.65 &0.986 &28.1 & 1276 & 151 &(3136)   &(-4.10)  &(2.87)   & (1.68) & -4.69\\
33 & 250 &6.96 &0.97 &40.2  & 1539 & 267 &(4085)   &(-3.96) &(2.82)    & (1.93)  & -4.40\\
34 & 153 &6.75 &0.975 &31.6 & 1359 & 215 &(3409)   &(-4.02) &(3.04)   & (1.81) & -4.58\\
\hline
\end{tabular}
\caption{High $\Gamma_{\rm e}$ mass-loss predictions for all three mass-range grids, with 
\teff\ kept constant at 50,000 K. 
The stellar parameters are given in Columns (2-7), providing 
the stellar mass, luminosity, Eddington factor, radius, escape velocity, and the effective escape 
velocity. Columns (8-11) give the wind properties: the terminal velocity, the mass-loss rate,
the wind-efficiency number \(\eta = \mdot \vinf/\lstar c\), and the wind acceleration parameter $\beta$. 
The last column provides the mass-loss rates as computed using the formula by Vink et al. (2000) with 
the computed terminal wind velocity as input.}
\label{tab:results}
\end{table*}

\subsection{Mass-loss predictions at high $\Gamma_{\rm{e}}$}
\label{sec_mdot}

Table~\ref{tab:results} lists our mass-loss predictions 
for all three considered mass ranges. Most columns are self-explanatory, but we note 
that the effective escape velocity $\vesc^{\rm eff}$ (7th column) is 
defined as $\sqrt{2GM(1-\gammae)/R}$. 
The predicted wind terminal velocities, mass-loss rates, wind efficiency numbers, and 
wind acceleration parameter $\beta$ are 
given in columns (8), (9), (10), and (11), respectively. 
For comparison column (12) lists the mass-loss values from the standard 
mass-loss recipe of Vink et al. (2000) where $\beta$ was held fixed at unity. 

\begin{figure}
\centerline{\psfig{file=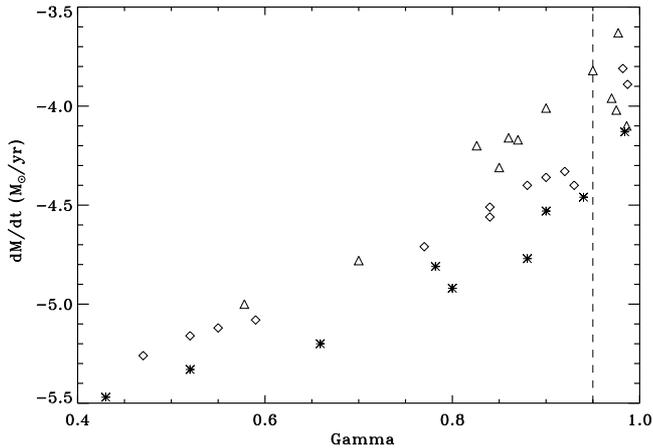, width = 9 cm}}
\caption{The predicted mass-loss rates versus $\Gamma_{\rm e}$ for models approaching 
the Eddington limit. Asterisks, diamonds, and triangles correspond to 
models of the respective mass ranges I, II, and III. Our model assumptions likely break down 
to the right of the vertical dashed line.} 
\label{f_mdot}
\end{figure}

\begin{figure}
\centerline{\psfig{file=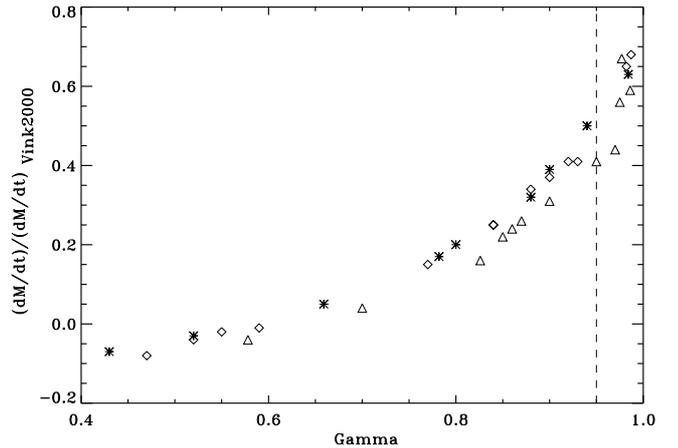, width = 9 cm}}
\caption{The logarithmic difference between the new $\Gamma_{\rm e}$ mass-loss predictions and the standard 
Vink et al. (2000) recipe for models approaching the Eddington limit. Symbols are the same as in Fig.\ref{f_mdot}.}
\label{f_diffmdotvink}
\end{figure}

The predicted mass-loss rates (column 9) are shown in 
Fig.~\ref{f_mdot}. Different symbols are used to identify the different mass ranges.
The figure shows that 
$\mdot$ increases with $\Gamma_{\rm e}$. This is 
in qualitative agreement with the luminosity dependence of the 
standard mass-loss recipe of Vink et al. (2000), derived from a 
set of models with \gammae\ $\la$ 0.4. 
Analogous to the results from the standard Vink et al. (2000) recipe, Fig.~\ref{f_mdot} suggests that 
there is an additional mass-loss dependence on mass, as for fixed \gammae\, the higher
mass stars have higher mass-loss rates. 
This finding confirms that mass-loss rates cannot solely be described 
by a dependence on luminosity or Eddington factor. This will be discussed 
further in Sect.~\ref{sec_gammae}. 

When comparing columns (9) and (12) from Table~\ref{tab:results}, it can be noted 
that our new high $\Gamma_{\rm e}$ mass-loss predictions tend to be 
larger than those determined using the standard Vink et al. (2000) recipe.
In order to quantify these differences, we divide the new mass-loss rates 
over those determined using the Vink et al. (2000) recipe (using the derived terminal wind velocities 
as input), and show the results 
in Fig.~\ref{f_diffmdotvink}. For the range $\Gamma_{\rm e}$ $\la$0.7, the 
differences are small. However, for values of 
$\Gamma_{\rm e}$ exceeding $\sim$0.7, the new and the old results diverge sharply. 
The maximum difference reaches a factor of five, which is similar in magnitude to what was  
reported previously for LBVs (Vink \& de Koter 2002) and WR stars 
(Vink \& de Koter 2005). We note that, although these prior results were based on 
global energy consistency with fixed $\vinf$/$\vesc$, where the velocity stratification was adopted, the reason 
for the differences revealed in Fig.~\ref{f_diffmdotvink} is that we probe a
different part of parameter space. 

\begin{figure}
\centerline{\psfig{file=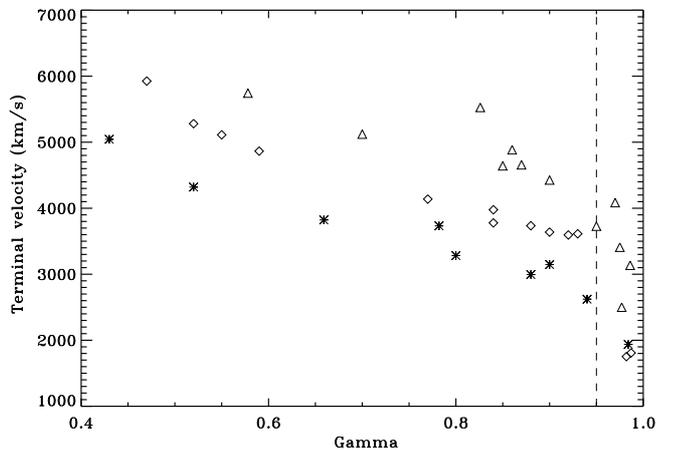, width = 9 cm}}
\caption{The predicted terminal wind velocities versus $\Gamma_{\rm e}$ for models approaching 
the Eddington limit. Symbols are the same as in Fig.\ref{f_mdot}.}
\label{f_vinf}
\end{figure}

We now turn our attention to the wind velocity structure. 
We first inspect the associated terminal wind velocity predictions. 
Figure~\ref{f_vinf} shows the behaviour of terminal wind velocity versus $\Gamma_{\rm e}$. 
The highest values are reached for the highest mass stars and 
exceed 5000 km/s. As expected $\vinf$ drops with $\Gamma_{\rm e}$. 
In the $\Gamma_{\rm e}$ range 0.4-0.95, the terminal wind velocity divided over 
the escape velocity is of the order 3-4, which is similar to the values for 
common O-type stars (Muijres et al. 2011b), where $\teff$ is in the range 30-40 kK, and 
the wind velocities are closer to 3000 km/s (see Sect.~\ref{sec_teff}). 

\begin{figure}
\centerline{\psfig{file=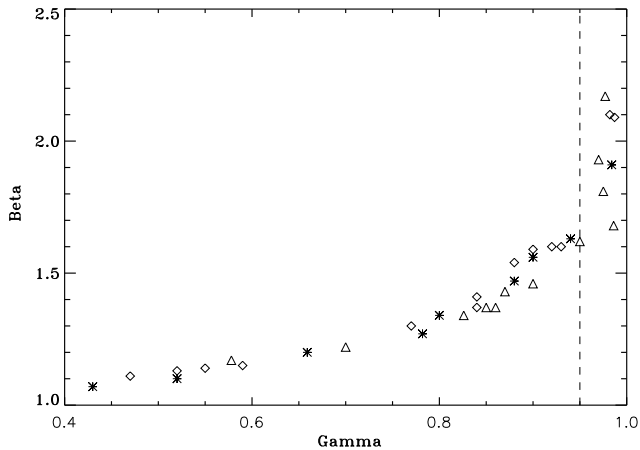, width = 9 cm}}
\caption{The predicted wind velocity structure parameter $\beta$ versus $\Gamma_{\rm e}$ 
for models approaching the Eddington limit. Symbols are the same as in Fig.\ref{f_mdot}.} 
\label{f_beta}
\end{figure}

We next turn our attention to the other wind velocity structure parameter, $\beta$, which 
describes how rapidly the wind accelerates. 
The predicted values of $\beta$ are depicted in Fig.~\ref{f_beta}, but  
$\beta$ does not show a significant dependence on stellar mass. 
For $\Gamma_{\rm e}$ up to 0.7, $\beta$ values are near unity, in accordance 
with the dynamical consistent models of Pauldrach et al. (1986), M\"uller \& Vink (2008), and Muijres et al. (2011b). 
However, when $\Gamma_{\rm e}$ exceeds 0.7 and approaches unity, $\beta$ steadily 
rises to values of about 1.7. These higher $\beta$ values are supposedly 
more commensurate in Wolf-Rayet stars (see e.g. Ignace et al. 2003), and it is reassuring 
to find that our models naturally predict this transition, without the use of any free parameter. 

{\em In all, our results suggest a natural extension from O-type mass loss to 
more extreme WR behaviour for increasing $\Gamma_{\rm e}$. 
An upturn in the $\dot{M}$ behaviour is found at $\Gamma_{\rm e}$ $\sim$0.7}.
Inspection of our models reveals a change from optically thin
to optically thick wind models at the position where we obtained 
the kink in the mass-loss versus $\Gamma$ relationship. 
Closer scrutiny of our model output also revealed that here the character of the Fe line driving 
changes. Whilst various ionization
stages of Fe contribute at all $\gammae$ models, we find that 
for the optically thin models at the low $\gammae$ end, just {\it one} ionization state of Fe 
dominates the relevant part of the wind driving domain (from just below the sonic point to 
about half the terminal velocity). By contrast,
for the optically thick models at the high $\gammae$ end, {\it two or more} ionization
stages of Fe contribute to the primary driving regime.

\subsection{ $\Gamma_{\rm{e}}$ dependence of mass loss}
\label{sec_gammae}

\begin{figure}
\centerline{\psfig{file=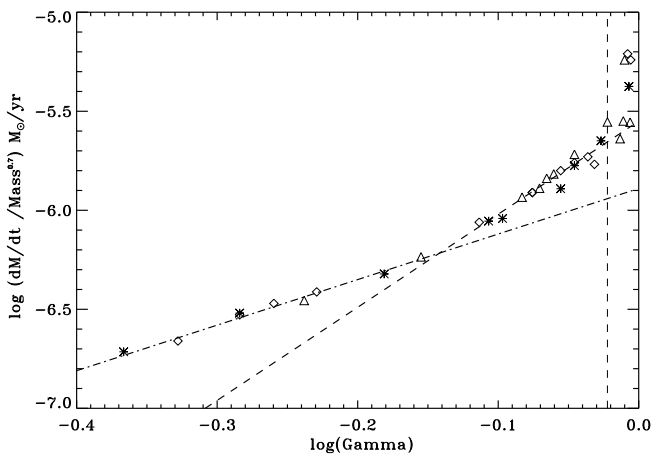, width = 9 cm}}
\caption{The predicted mass-loss rates divided by $M^{0.7}$ versus $\Gamma_{\rm e}$
for models approaching the Eddington limit. The dashed-dotted line represents
the best linear fit for the range 0.4 $<$ \gammae $<$ 0.7. The dashed line 
represents the higher 0.7 $<$ \gammae $<$ 0.95 range. Symbols are the same as in Fig.\ref{f_mdot}.} 
\label{f_mdotmass}
\end{figure}

In order to determine the dependence of the mass-loss rate on $\Gamma_{\rm e}$, 
we could simply fit the datapoints to a power law:
\begin{equation}
\label{eq:fit}
	\mdot \propto \gammae^{p}	
\end{equation}
Using the semi-empirical approach Vink (2006) found $p$ to be equal to $\sim$5, and our dynamically consistent 
results provide the same slope here. However, in order to also take 
the {\it mass} dependence into account we divide the mass-loss rates 
by $M^{q}$, and show the results in Fig.~\ref{f_mdotmass}. We 
fit the data with the following power-law

\begin{equation}
\label{eq:fit1}
	\mdot \propto \mstar^{q} \gammae^{p}	
\end{equation}
Below  \gammae\ $\la$ 0.95, Fig.~\ref{f_mdotmass} shows two 
mass-loss regimes, divided by a boundary at \gammae\ $\sim$ 0.7. 
This is not only the point where the slope of the mass-loss versus
$\Gamma$ relation changes, but also where 
the wind efficiency parameter $\eta$ surpasses the single scattering limit (see below). 
Upon further inspection of our models, we find that 
as long as \gammae\ $\la$ 0.7 the winds 
are optically thin, implying that the sonic point of the outflowing material 
lies outside the photosphere, whilst the winds become optically thick -- with the photosphere moving 
outside of the sonic point -- for \gammae\ values above $\ga$0.7. 

We derive two independent mass-loss relationships 
for the two separate $\Gamma_{\rm e}$ regimes.\\

\noindent
For 0.4 $<$ \gammae $<$ 0.7 we find

\begin{eqnarray}
\label{eq:mlrB1}  \nonumber
\dot{M} \propto M^{0.68} \gammae^{2.2}
\end{eqnarray}

\noindent
For 0.7 $<$ \gammae $<$ 0.95 we determine that

\begin{eqnarray}
\label{eq:mlrB}  \nonumber
\dot{M} \propto M^{0.78} \gammae^{4.77}
\end{eqnarray}

We first note that we intentionally do not provide equations here, as 
we expect the absolute values of these mass-loss predictions 
for the high $\Gamma$ models to be underpredicted. 
In Sect. \ref{s_emp}, we see 
that our predicted wind terminal velocities are  
a factor 2-4 higher than found empirically. 
If we had employed our older semi-empirical
approach, and assumed empirical -- i.e. a factor 2-4 lower -- 
terminal wind velocities, we would have predicted higher mass-loss rates. 
Indeed, the pilot study of Vink (2006) that was based on the 
semi-empirical approach provided such higher mass-loss rates.  

Secondly, we note that the exact value of the $\gammae$ transition value 
is model-dependent. We have already discussed our potential model deficiencies
in the deepest layers in Sect. \ref{sec_applic}, but we should note that 
several other factors might also play a role;  
in particular, higher mass-loss rates -- as would be obtained from our semi-empirical 
approach -- would shift the kink to lower $\gammae$ values. 
We note that the latter could occur in the case the wind is clumped. 
As long as porosity effects are small, wind clumping 
is expected to increase the momentum transfer in our MC models (Muijres et al. 2011a). 
Although future dynamical-consistent modelling of clumped winds is required to test this,  
wind clumping could potentially increase our predicted mass-loss rates, and subsequently 
decrease the $\gammae$ value of the kink. 

The above mass-loss relationships can easily be transformed using 
Eq.~(\ref{eq_gammae}), leading to the entirely analogous mass-loss relationships 
$\dot{M} \propto L^{0.68} \gammae^{1.52}$ and Eq.~(\ref{eq:mlrB})
to $\dot{M} \propto L^{0.78} \gammae^{3.99}$.
Interestingly, if one subsequently applies a mass-luminosity relationship 
for classical (He-rich) WR stars of Maeder \& Meynet (1987) or for very massive H-rich 
stars such as that of Yungelson et al. (2008), with
$L \propto M^{1.34}$ for both cases, it follows that $\dot{M} \propto M^{2.4}$. This appears 
to be in good accord with the radio mass-loss rate relation $\dot{M} \propto M^{2.3}$ for 
classical WR stars with measured masses from binaries by Abbott et al. (1986). 
It also agrees with the $\dot{M}$ versus stellar mass relationship 
of $\dot{M} \propto M^{2.5}$ that has been applied  
in WR evolution models by Langer (1989).

\begin{figure}
\centerline{\psfig{file=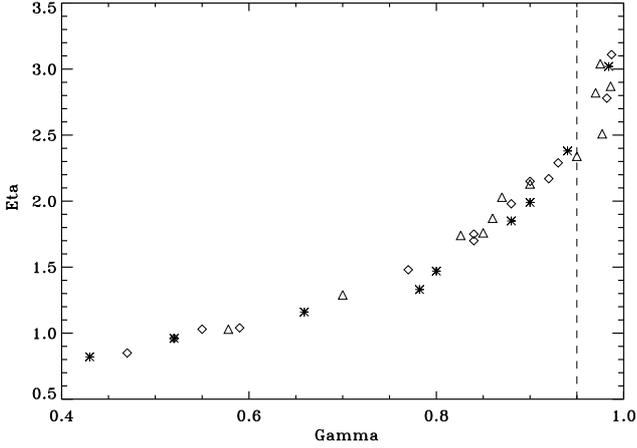, width = 9 cm}}
\caption{The predicted wind efficiency number $\eta$ versus $\Gamma_{\rm e}$ for models approaching 
the Eddington limit. Symbols are the same as in Fig.\ref{f_mdot}.} 
\label{f_eta}
\end{figure}

\subsection{Increased wind efficiency close to the Eddington limit?}
\label{sec_maxml}

In order to learn whether radiation-driven mass-loss rates 
continue to increase with increasing \gammae\ or reach a maximum in 
\mdot\ instead, it is insightful to consider the wind efficiency 
parameter $\eta = \mdot \vinf / (L/c)$. 
We show the predicted values of $\eta$ in Fig.~\ref{f_eta}. 
As the symbols denote different mass ranges, the small scatter on the datapoints 
shows that $\eta$ is not very sensitive to stellar mass. 
At values of $\Gamma_{\rm e} \sim$ 0.5 we find wind efficiency 
numbers $\eta$ of order 1, in accordance with standard Vink et al. (2000) models. 
However, when $\Gamma_{\rm e}$ approaches unity, 
$\eta$ rises in a curved manner to values as high as $\eta$ $\simeq$ 2.5. 
Such high $\eta$ values are more commensurate with Wolf-Rayet winds 
than with common O star winds, and these results thus confirm a natural extension from 
common O-type mass loss to more extreme WR behaviour.
In Sect.~\ref{s_emp}, we find that our predicted wind-terminal velocities 
are higher than the empirical values. As we note that an overprediction of the wind velocity
is likely offset by a mass-loss rate underprediction by a similar amount, we argue that
the combination of these two quantities, i.e. their product constituting 
the $\eta$ parameter, might be less affected by model deficiencies than 
either of these quantities would be individually.

The maximum mass loss in our
models up to \gammae\ = 0.95 is  log \(\mdot_{\rm{max}}\) = -3.8. 
This is the mass-loss rate that is retrieved for the most extreme models in our grid. 
Owocki et al. (2004) have investigated the mass loss of stars that formally exceed their Eddington
limit and show that the expected mass loss falls well below the values 
required to account for the mass that is lost during LBV giant eruptions, such as that of \(\eta\) Carinae in the 1840s. 
Interestingly, they introduce a porosity-moderated {\em continuum} driven mass loss that might account for 
the huge mass-loss rates associated with LBV eruptions (which may be of the order of 1$\msun$/yr).

\subsection{Effect of \teff\ on high $\Gamma_{\rm{e}}$ models}
\label{sec_teff}

\begin{figure}
\centerline{\psfig{file=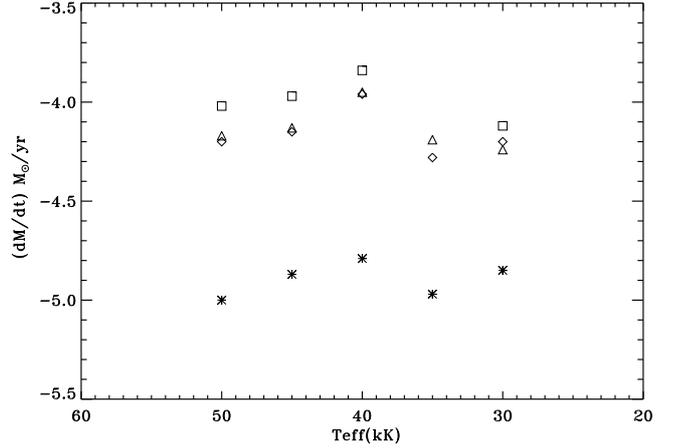, width = 9 cm}}
\caption{The predicted mass-loss rates versus effective temperatures 
for several values of $\Gamma_{\rm e}$, with from top to bottom 
\gammae\ equal to 0.90 (model 29; open square), 0.87 (model 28;open triangle), 0.83 (model 25;open diamond), and 0.58 (model 23; asterisk), respectively.} 
\label{f_teff}
\end{figure}

\begin{figure}
\centerline{\psfig{file=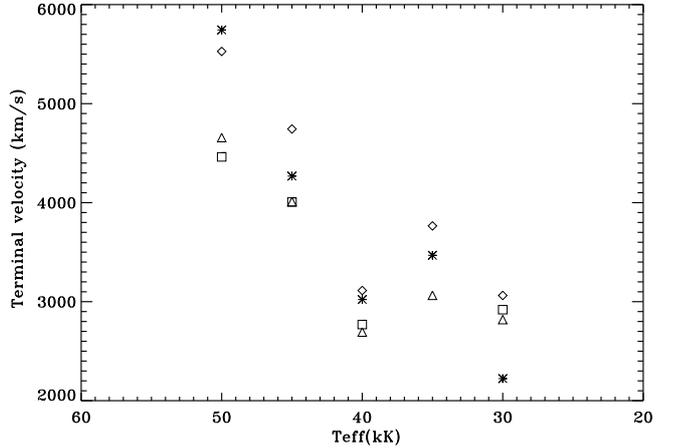, width = 9 cm}}
\caption{The predicted terminal velocities versus effective temperatures 
for several values of $\Gamma_{\rm e}$, with \gammae\ equal to 0.90 (model 29; open square), 0.87 (model 28;open triangle), 0.83 (model 25;open diamond), and 0.58 (model 23; asterisk), respectively.}
\label{f_vinfteff}
\end{figure}

To establish whether there is an additional temperature 
dependence on $\dot{M}$, we varied 
\teff\ over the range 50-30\,kK for selected \gammae\ models, with mass-loss 
predictions presented in Fig.~\ref{f_teff} and terminal wind velocities shown 
in Fig.~\ref{f_vinfteff}. As we wish to stay above the temperature of the 
bi-stability jump (which starts at \teff\ values below $\sim$27.5\,kK; see Vink et al. 2000), 
we restrict our \teff\ range to a minimum value of 30 kK.
We find that for all \gammae\ values \mdot\ is not a strong function of temperature.
In terms of the terminal velocity dependence, Fig.~\ref{f_vinfteff} shows 
a rather steep dependence on temperature, with $\varv_{\infty}$ dropping by a factor two. 
This is merely a reflection of the escape velocity dropping by a similar 
factor of two over the temperature range under consideration.

\subsection{Effect of the helium abundance on high $\Gamma_{\rm{e}}$ models}
\label{sec_he}

\begin{table*}
\centering
\begin{tabular}{rrrlllll}
\hline
\hline\\[-6pt]
model & $\mstar_{\rm old}$ & $\mstar_{\rm new}$ & $\log L$ & \gammae & \vinf & $\log \mdot$ &\(\beta\)\\

number & [\msun] & [\msun] & [\lsun] &   & [\kmsec] &[\msunyr] &  \\
\hline
\\

2He & 60  &35.8 & 6.0    &0.43 &4552  &-5.36 & 1.12\\
5He & 40  &23.1 & 6.08   &0.80 &3141  &-4.86 & 1.42\\
10He & 85 &49.0 & 6.25   &0.55 &4567  &-5.04 & 1.21\\
14He & 85  &49.0  & 6.39 &0.77 &4144  &-4.69 & 1.40\\
24He & 120 &69.5  & 6.50 &0.70 &4800  &-4.70 & 1.32\\
26He & 180 &104.1 & 6.76 &0.85 &4759  &-4.28 & 1.53\\
29He & 275 &159.5 & 6.97 &0.90 &4958  &-3.99 & 1.70\\
30He & 300 &175.0 &7.03  &0.95 &4934  &-3.88 & 1.75\\[2pt]
\hline
\end{tabular}
\caption{Helium-enriched mass-loss predictions. All parameters that
are not listed are the same as in Table~\ref{tab:results}, and 
masses have been lowered to keep $\gammae$ fixed.
}
\label{tab:resultshe}
\end{table*}

\begin{figure}
\centerline{\psfig{file=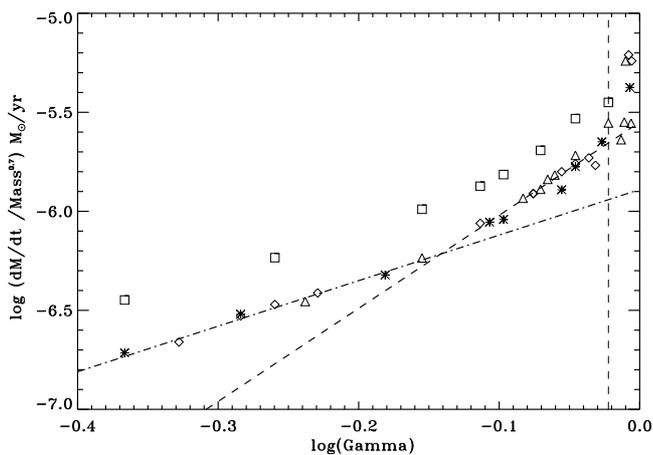, width = 9 cm}}
\caption{Same as Fig.~\ref{f_mdotmass} but for He models from Table~\ref{tab:resultshe} added 
as open straight squares. The He models form a straight line above the H-rich relationship.
}
\label{f_mdot_mass_he}
\end{figure}

To establish the existence of a potential helium dependence on $\dot{M}$, 
we computed additional models across the entire \gammae\ region, 
setting the hydrogen abundance to zero and increasing the helium abundance accordingly. 
The results are listed in Table~\ref{tab:resultshe} and shown 
in Fig.~\ref{f_mdot_mass_he}. The mass-loss rates are similar 
to those of H-rich models for objects with the same \gammae\ (see Table~\ref{tab:results}). 
This is not too surprising given that the indirect effects of different continuum energy distributions
for H-rich versus H-poor are rather subtle (Vink \& de Koter 2002). 
However, when the He-rich results are plotted in Fig.~\ref{f_mdot_mass_he} they lie above the H-rich models. For equal 
luminosity and \gammae\ the masses of the He-rich models are lower since \gammae\ is a function of
the chemical composition through $\sigma_{\rm e}$ (see Eq.~\ref{eq_gammae}); $\sigma_{\rm e}$ is lower for He-rich 
models, therefore the mass must be lowered to keep \gammae\ constant. Similar to the H-rich models, 
there appears to be an upturn in the mass loss vs. \gammae\ dependence for models 
at about \gammae\ $\sim$ 0.7. 

With respect to the terminal velocity and $\beta$-dependence, we do not find any significant 
differences between H-rich and He-rich models (see Table~\ref{tab:results} versus Table~\ref{tab:resultshe}).

\section{Spectral morphology: the characteristic He 4686 Angstrom line}
\label{s_morph}

In the previous section, we provided evidence for a natural 
transition in the mass-loss-$\Gamma_{\rm e}$ exponent, as well as in the 
velocity parameter $\beta$ and wind-efficiency $\eta$ from moderate $\Gamma_{\rm e}$ 
``optically thin wind'' cases to ``optically thick wind'' cases for objects that find themselves above 
$\Gamma_{\rm e}$ $\ga$ 0.7. We have inspected our models and confirmed that 
for $\Gamma_{\rm e}$ $\la$ 0.7 the sonic velocity is reached outside the 
photosphere, whilst the stars form a pseudo-photosphere for $\ga$ 0.7. 

\begin{figure}
\centerline{\psfig{file=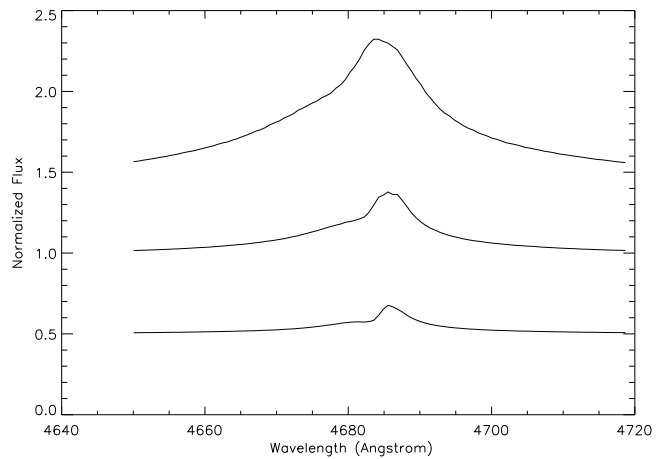, width = 9 cm}}
\caption{The predicted normalized He\,{\sc ii}\,$\lambda 4686$ flux versus wavelength
for three values of $\Gamma_{\rm e}$, with from top to bottom 
\gammae\ equal to 0.93 (model 20; 90\msun; EW $\sim$$-$20\AA), 0.84 (model 15; 100\msun; 
EW $\sim$$-$7\AA), 0.70 (model 24, 120\msun; EW $\sim$$-$2\AA), respectively.
Wind clumping has not been taken into account here.} 
\label{f_he}
\end{figure}

We expect that the occurrence of a pseudo-photosphere has a consequence for 
the spectral morphology of the stars in question.
We might suspect that the transition $\Gamma_{\rm e}=$ 0.7 is the point where 
the spectral morphology of normal O stars changes from the common O and Of-types 
into a WN-type spectrum. The spectral sequence involving 
the Of/WN stars has a long history (e.g. Conti 1976, Walborn et al. 1992, 
de Koter et al. 1997, Crowther \& Dessart 1998) but it still has to be placed into 
a theoretical context. Figure~\ref{f_he} shows a sequence for the predicted 
He {\sc ii} 4686\AA\ lines for three gradually increasing values of $\Gamma_{\rm e}$: 
0.70 (model 24), 0.84 (model 15), and 0.93 (model 20), respectively. These
models have been selected to be objects with a constant luminosity of \logl\ $=$ $6.5$, and we  
simply lowered the mass from 120\msun\, to 100\msun, to 90\msun.
It is insightful to note that, although the first spectrum below 
the transition $\Gamma_{\rm e}$ already shows some emission -- characteristic of Of stars -- the line-flux 
is rather modest in comparison to what is found for the next two cases with $\Gamma_{\rm e}$ values 
exceeding the critical value of 0.7.
These objects show very strong and broad He {\sc ii} 4686\AA\ emission lines that are 
more characteristic of Of/WN or ``slash'' stars, progressing towards the 
Wolf-Rayet stars of the nitrogen sequence (WN). 

These models thus indicate that the observed spectral transition 
from Of to WN corresponds to a transition from relatively low \gammae\ to high \gammae\ values (and larger $\beta$) for WN stars.   
This assertion is based not only on the higher predicted mass-loss rates themselves, but also on the finding that 
the mass-loss behaviour (as a function of \gammae\ ) changes at $\Gamma_{\rm e}$ $=$ 0.7.
We note that the increasing He{\sc ii} 4686\AA\ equivalent width (EW) 
amounts to EW values of $-$2, $-$7, $-$20$\AA$, respectively (for these unclumped models).

\section{Discussion}
\label{disc}

\subsection{Comparison with empirical mass-loss rates}
\label{s_emp}

Comparing our new mass-loss predictions against observed mass-loss rates 
is a non-trivial undertaking, as high $\Gamma_{\rm e}$ objects are scarce. 
The largest sample of such potentially high $\Gamma_{\rm e}$ objects 
that involves state-of-the-art modelling analysis is probably that 
of the Arches cluster by Martins et al. (2008). They provided stellar and wind 
properties (accounting for wind clumping) of 28 of its brightest members 
from $K$-band spectroscopy. Roughly half of their sample comprises O4-O6 supergiants whilst the 
other half includes H-rich WN7-9 stars. 

It is not possible to quote direct mass-loss predictions, because 
the Martins et al. analysis did not yield object masses. 
However, on the basis of the high stellar luminosities, with $\logl$ up to 6.3, 
these objects were suggested to be consistent with initial masses of up to 
$\sim$120$\msun$. 
For the O4-6 supergiant population, luminosity 
values are in the range $\logl$ $=$ 5.75--6.05, consistent with initial masses 
$M \simeq 55-95 \msun$. For this mass and luminosity range, 
$\Gamma_{\rm e}$ $\simeq$ 0.2 -- comfortably within our low-$\Gamma_{\rm e}$ 
regime. Assuming the current mass of these objects is about the same as their 
initial mass, our mass-loss formula yields values of $\log \mdot$ $\simeq$ $-$6.1, which 
is in reasonable agreement with the lower end of the Martins et al. mass-loss 
rates for their O4-O6 I objects. 

The second group of Martins et al. objects 
comprise the WN7-9 objects. If we again assume their current masses can 
directly be inferred from the observed luminosities, we find $\Gamma_{\rm e}$ 
$\simeq$ 0.4 and mass-loss rates $\log \mdot \simeq -5.3$.
Even if the helium abundances of these objects are increased, these properties 
would not result in a pronounced emission profile of the 
He\,{\sc ii}\,$\lambda 4686$ line, i.e. a profile shape that is typical 
of late-WN stars. For this to happen the mass-loss rates need to be higher
by at least a factor of a few. This seems to require high $\Gamma$. In the framework 
of our models, this could be achieved by lowering the mass. However as the non-electron 
contribution in our models is not self-consistently treated in the high $\Gamma_{\rm e}$ 
regime, we refrain from providing quantitative assessments. 

As the O4-O6 supergiants from the Martins et al. (2008) analysis have 
effective temperatures in the range 32-40 kK, we expect $\varv_{\infty}$
to fall in the range 2500-3500 \kmsec\ (see Fig.~\ref{f_vinfteff}), which reasonably 
agrees with the upper end of the Arches O4-O6 supergiant stars. However, for the 
late WN stars, the terminal velocities presented by Martins et al. drop to values as low as 800-1600 
\kmsec, which is a factor 2-4 lower than we predict. We identify a number of 
possible reasons for this discrepancy. One option could be 
that the K-band spectral fits of Martins et al. (2008) yield terminal velocities that are 
too low (note that no ultra violet P Cygni blue edges are available for these
obscured objects), but a more plausible reason is that as a result of 
our modelling assumptions (no rotation, smooth winds, etc.), we overpredict the wind 
terminal velocity. As the overprediction in terminal velocity is a factor 2-4, this would naturally 
translate in a mass-loss underestimate of a factor 3-5.
 
Although the effects of rotation on our mass-loss predictions are beyond the scope of this 
investigation, rotation may become 
relevant once one wishes to compare the non-rotating predictions to observations of 
objects that might rotate at a relevant rate. In particular, given that 
Gr\"afener et al. (2011) discuss the possibility that the 
WNh stars in the Arches cluster may evolve close to being chemically homogeneous.
To first order, one might expect the effect 
of rotation to lower the effective gravity; i.e. one could replace the mass by the 
effective mass $M(1-\Gamma)(1-\Omega^2)$, in which case one would anticipate $\dot{M}$ to scale as
$\dot{M}(\Omega) \propto \dot{M}(0) (1-\Omega^2)^{1-\frac{1}{\alpha'}}$. In a similar vein, one 
might expect the terminal velocity to scale with the escape velocity, i.e. 
$v_{\infty} \propto v_{\infty}(0) \sqrt{(1-\Omega^2)}$\footnote{The issue of rotation in radiation-driven wind
is highly complex. All existing studies (e.g. Friend \& Abbott 1986, Bjorkman \& Cassinelli 1993, Petrenz \& Puls 2000, Pelupessy et al. 2000, 
Cure \& Rial 2004, Madura et al. 2007) have 
only treated part of the problem, but not tackled the combined 
multi-dimensional, high $\Omega$ ($\Omega \ga 0.75$), and high $\Gamma$ aspects.} 
(see e.g. Gayley  2000; Puls et al. 
2008 and references therein). In Sect.~\ref{sec_he}, 
we computed some test models in which we lowered the stellar 
mass, hence the effective gravity, so as to keep \gammae\ 
constant (in order to study a potential helium dependence on $\dot{M}$). 
These models indeed showed higher mass-loss rates 
for lower effective gravity, but can we quantify the effective-gravity 
effect? 

If we were to attribute the offset between the observed and predicted wind-terminal velocity 
of a factor $\sim$2 to stellar rotation, we should have an $\Omega$ of $\sim$0.85 
to ``match'' theory to observations. 
In that case, the mass-loss rate would be enhanced by a factor of $2.5-4$ for $\alpha'$ values 
in the range 0.5-0.6. Nevertheless, this could significantly change for lower temperatures.
These high $\Omega$ values of the order of 0.85 seem to be rather high 
when we consider recent stellar evolution models (Brott et al. 2011; 
Friedrich et al. in prep.), but they are not out of line with respect 
to spectral observations of LBVs (Groh et al. 2006), objects 
that are presumably in close proximity to the Eddington limit.

In all, we interpret our high wind terminal velocities 
as a sign that some physics is missing in our high $\Gamma$ models. 
Therefore, we refrain from using our dynamically determined mass-loss rates in 
a quantitative way at the optically-thick high $\Gamma$ end. 
For normal O-type stars, for low and moderate $\gammae$ values, we achieve much better agreement between 
observed and predicted wind terminal velocities (Muijres et al. 2011b), 
and for this regime we have a much higher confidence in the accuracy of the absolute mass-loss rates, 
as long as O-type winds are not extremely porous, in which case 
mass-loss rates could drop significantly (Muijres et al. 2011a).

\subsection{Comparison to other models}
\label{s_theory}

\subsubsection{Comparison to CAK and other O-type star mass-loss models} 

We now wish to compare our results with previous model predictions. 
In this paper, we have investigated the mass-loss behaviour at high \gammae\ for 
an extensive grid of models, and we revealed the existence of two mass-loss regimes. 
Moreover, we have found that mass-loss rates are dependent on both \gammae\ and stellar mass
(or stellar luminosity) and that the shape of these dependencies is well-described by a power law.

We remind the reader of classical CAK theory, where 
the mass-loss rate is found to be proportional to

\begin{equation}
\label{eq_betalaw}
\mdot~\propto~L~\left(\frac{\gammae}{1~-~\gammae}\right)^\frac{1~-~\alpha}{\alpha}
\label{eq_cak}
\end{equation}
where $\alpha$ is a force multiplier parameter expressing the importance of optically thin
lines to the total ensemble of lines. It is generally found that $\alpha$ is $\sim$2/3 for 
galactic O-type stars (Puls et al. 2008) and assumed to be constant throughout the atmosphere. 
In reality, however, $\alpha$ is depth-dependent (Vink 2000, Kudritzki 2002, Gr\"afener \& Hamann 2005, 
Muijres et al. 2011b), which is captured better by an alternative representation of the 
line acceleration (M\"uller \& Vink 2008).
Nevertheless, the classical CAK formalism -- as described by 
Eq.~\ref{eq_cak} -- already shows a dependence on both $L$ and $\gammae$, and one could 
rewrite this mass-loss dependence as a function of $M$ and $\gammae$, using a 
mass-luminosity relation. 

In the standard Vink et al. (2000) mass-loss parametrization  
$\mdot$ $\propto$ $L^{2.2}$ $M^{-1.3}$ $(\vinf/\vesc)^{-1.2}$, which
can be reorganized to $\mdot$ $\propto$ $L^{1.2}$ $\gammae^{0.7}$. 
This is the type of mass-loss parametrization that is currently employed in   
modern evolutionary computations (see e.g Meynet \& Maeder 2003, Palacios et al 2005, 
Limongi \& Chieffi 2006, Eldridge \& Vink 2006, Vink et al.
2010, and Brott et al. 2011). 

In this paper, we obtain much steeper \mdot\ vs. $\Gamma$ dependencies, in 
agreement with our previous models for constant-luminosity 
LBVs (Vink \& de Koter 2002, Smith et al. 2004).
For the ``low'' $\gammae$ range considered here we obtained ``modest'' 
dependencies of $\mdot$ $\propto$ $L^{0.7}$ $\gammae^{1.5}$ , but 
for the ``high'' $\gammae$ regime, we found $\mdot$ $\propto$ $L^{0.78}$ $\gammae^{3.99}$
which involves a much steeper dependence on $\gammae$ than any
CAK-type mass-loss relationship provides. 
We note that such a steep dependence on the Eddington
limit agrees with radiation-driven wind models of Vink (2006) and Gr\"afener \& Hamann (2008), whilst 
Gr\"afener et al. (in prep.) also provide empirical evidence for such a strong mass-loss dependence 
on the Eddington parameter. 
 
We emphasize that, through the use of the Vink et al. (2000) theoretical mass-loss recipe,
most current stellar models already include the effect of positive mass-loss feedback 
(contrary to recent claims by Smith \& Conti 2008). This effect 
describes how the mass-loss rate increases with the Eddington parameter. 
However, as we here obtain much steeper \mdot\ vs. $\Gamma$ dependencies, 
it is likely that the mass-loss feedback 
effect that is currently employed in the stellar evolution models may not be 
sufficient for certain areas of the Hertzsprung-Russell diagram. 
We thus concur with the notion of Smith \& Conti (2008) that new stellar evolution 
computations that take this effect into account properly are desirable

\subsubsection{Comparison to alternative Wolf-Rayet mass-loss models}

We also compare our models to the optically thick wind models for Wolf-Rayet stars, such
as the critical-point analysis by Nugis \& Lamers (2002) and  the hydrodynamical 
model atmosphere analysis of Gr\"afener \& Hamann (2008).
As there is a significant qualitative difference between our Monte Carlo approach and 
the optically thick wind approaches, a meaningful quantitative comparison is a non-trivial 
undertaking (see section~\ref{sec_applic}). 

First, we quantitatively compare our $\dot{M}$ versus $\Gamma_{\rm e}$
dependence to the WNL star mass-loss dependence suggested by
Gr\"afener \& Hamann (2008). For the models in our grid at $\teff$ = 50 kK, we find 
very good agreement with the Gr\"afener \& Hamann (2008) mass-loss rates and 
also find that the power-law slope of our
dependence is very similar. However, the onset of WR-type behaviour
occurs earlier, i.e. for lower $\Gamma_{\rm e}$, in the models by
Gr\"afener \& Hamann.

In Sect.~\ref{sec_pspace}, we discussed the possibility of such a
shift in $\Gamma_{\rm e}$, because the actual Eddington parameter
$\Gamma$ is expected to be affected by free-free and bound-free contributions 
and peaks in the iron opacity. By comparison with OPAL opacity tables (Iglesias \& Rogers
1996), we estimate an increase of $\Gamma$ by $\sim 20\%$ in the region
of the sonic point, assuming the location of the sonic point remains unaffected.
This value corresponds roughly to a $\sim
25\%$ shift in $\Gamma_{\rm e}$ between our relation and that by Gr\"afener \&
Hamann for typical parameters of Galactic WNL stars ($T_{\rm eff}=45$\,kK, $\log(L/\Lsun) =6.3$). 
This could be considered a maximum shift as we may slightly overestimate the line force 
near the sonic point by applying the Sobolev approximation (Pauldrach et al. 1986). However, a change in $\Gamma$
affects the atmospheric structure and therefore the location of the sonic point, consequently the effect on 
$\dot{M}$ is hard to establish. 
  
The Gr\"afener \& Hamann mass-loss rates also display a strong temperature dependence,
with $\dot{M} \propto T_{\rm eff}^{-3.5}$. The actual size of the shift
in $\Gamma_{\rm e}$ is thus strongly dependent on the specific stellar parameters.
Our Monte Carlo models suggest a much smoother
dependence on $\teff$ (see Fig.~\ref{f_teff}) 
as long as we stay above the location of the predicted bi-stability jump, where the mass-loss properties 
jump drastically (Vink et al. 1999, Pauldrach \& Puls 1990).
  
What we wish to emphasize is that both modelling approaches show an $\mdot$ versus 
$\gammae$ dependence that is {\it much} stronger than any additional mass or 
luminosity dependence. Where the two distinct mass-loss prescriptions differ
is in the treatment of wind clumping, as well as the value of $\Gamma$ for 
the onset of WR-type mass loss behaviour. 
The exact location of this
transition is of paramount importance for the evolution of the most massive stars.
Ultimately, this should be
testable with comparisons to observational data when sufficient
objects are available in the appropriate $\Gamma_{\rm e}$ range. 
This will be a crucial aim of the VLT Flames Tarantula Survey (
Evans et al. 2011, Bestenlehner et al. in prep.). 

\section{Summary}
\label{s_sum}

We presented mass-loss predictions from Monte Carlo radiative 
transfer models for very massive stars in the mass range 40-300\msun\ and 
with Eddington factors \gammae\ in the range 0.4--1.0. 
An important outcome is that when winds become optically thick the spectral and 
mass-loss properties change.
This transitional behaviour can be summarized as follows

\begin{itemize}

\item{We find a {\it transition} from common O-type stars 
to more extreme Wolf-Rayet behaviour when \gammae\ exceeds a critical value.}

\item{The way in which the mass-loss rate depends on \gammae\ 
in the 
 range 0.4 $\la$ $\Gamma_{\rm e}$ $\la$ 0.7 is
\mdot\ $\propto$ $\mstar^{0.68} \Gamma_{\rm e}^{2.2}$, where rates
are found to be consistent with the standard Vink et al. (2000) mass-loss rates.}

\item{At $\gammae \simeq 0.7$ the \mdot\ dependence shows a ``kink''; i.e. the
      slope is steeper for objects closer to the Eddington limit. Here the
      slope becomes \mdot\ $\propto$ $\mstar^{0.78} \Gamma_{\rm e}^{4.77}$. This slope 
      agrees with WNL models by Gr\"{a}fener \& Hamann (2008).}
      
\item{When $\Gamma_{\rm e}$ approaches unity, the wind efficiency number {\it $\eta$ rises in a 
curved manner to values as high as $\eta$ $\simeq$2.5}. 
Such high $\eta$ values are more commensurate with Wolf-Rayet winds 
than with common O stars winds, and these results thus confirm a natural extension from 
common O-type mass loss to more extreme WR behaviour.}

\item{This transitional behaviour is also found in terms of the wind acceleration
parameter {\it $\beta$, which naturally reaches values as high as 1.5}}

\item{The spectral morphology of the He {\sc ii} line at 
4686\AA\ changes gradually as a function of $\Gamma_{\rm e}$. This links the spectral sequence 
O-Of-Of/WN-WN to a transition of optically thin to optically thick winds.}

\item{The mass-loss rate is found to be only modestly dependent on the effective temperature
 for the range of 30 to 50 kK.}

\item{Last but not least, we highlight the fact that for fixed \gammae\ the He abundance only 
has a minor effect on the predicted rate of mass loss (cf. Vink \& de Koter 2002 for LBVs).
This contradicts how O-type, to LBV-type, to WR-type mass-loss transitions 
are employed in massive star evolution models. We thus call for fundamental changes in the way 
mass loss is included in stellar evolution models for objects in close proximity 
to the Eddington limit.}

\end{itemize}


\begin{acknowledgements}
We would like to thank the anonymous referee for providing 
constructive comments.  
\end{acknowledgements}

\end{document}